\documentclass[%
 reprint,
 superscriptaddress,
 showpacs,preprintnumbers,
 amsmath,amssymb,
 aps,
 pra,
 longbibliography,
lengthcheck,%
]{revtex4-1}

\usepackage{graphicx}
\usepackage{dcolumn}
\usepackage{bm}
\usepackage{color}
\usepackage{ulem}
\usepackage{hyperref}

\begin{document}

\preprint{APS/123-QED}

\title{
  Temperature-dependent magnetic anisotropy from directional-dependent interactions 
}

\author{Hiroaki Ishizuka}
\affiliation{
Department of Applied Physics, The University of Tokyo, Bunkyo, Tokyo, 113-8656, JAPAN 
}

\date{\today}

\begin{abstract}
  Magnetic anisotropy of spin models with directional-dependent interactions in the high-temperature paramagnetic phase is theoretically studied. Using a high temperature expansion, we show that the Ising type directional-dependent interaction gives rise to magnetic anisotropy which depends on the temperature as $\propto T^{-5}$. This phenomenon arises from the anisotropic exchange interaction, and is distinct from the orbital effect, such as van Vleck susceptibility. It is shown that while the quadratic term in the magnetization favors to point the spins along the bond, the fourth order term in magnetization prefers to point spins to the perpendicular direction. The theory is applied to the Heisenberg-Kitaev model on the honeycomb lattice and a cubic lattice model that is potentially relevant to perovskite iridates. We show that, in these models, the anisotropic terms in quadratic order cancels out, and the leading order for the magnetic anisotropy arises from the fourth order contribution. The result shows that the anisotropy from the directional-dependent interaction gives rise to $\langle100\rangle$ magnetic anisotropy. These results are potentially relevant to heavy transition metal oxides such as iridates. Experimental observation of the magnitude of anisotropic interactions using magnetic torque measurement is also discussed.
\end{abstract}

\pacs{
}

\maketitle

\section{Introduction}

The effect of spin-orbit interactions in magnets are of considerable interest, for its rich physics in magnetism the interaction gives rise to. An important effect of the spin-orbit interaction in magnets is that the interaction lifts the orbital degeneracy, potentially realizing the pure spin systems in solid state materials free from orbital physics. A consequence of such orbital deneneracy lifting is that the spin-orbit interaction bring about anisotropic interactions that depends on the bond. While such interaction is expected to be relatively small in the $3d$ transition metal oxides, the interaction can be prominently enhanced in heavy transition metal compounds, such as $4d$ and $5d$ systems, and also in rare-earth compounds. Indeed, it was pointed out that, in a certain setup, the directional-dependent interaction can become the only exchange interaction between the local moments~\cite{Jackeli2009}. Therefore, studies on the existance and effects of anisotropic exchange interactions on the magnetic ground state is a key to understand the physics of the heavy ion magnets, and have been investigated widely, for example in Refs.~\onlinecite{Jackeli2009,Chen2008,Chen2010,Bhattacharjee2012,Savary2012,Ishizuka2014,Rau2014,Yamaji2014,Chaloupka2016,Ohkubo2016}.

Moreover, it have been revealed that the anisotropic interactions give rise to rich physics. One of the well studied example of such anisotropic exchange interaction is the directional-dependent interaction --- the Ising type interaction which the spin component depends on the bond. In the ground state, the directional-dependent interaction gives rise to rich physics, engenders quantum spin liquid state~\cite{Kitaev2006} and bring about the magnetic anisotropy from fluctuations~\cite{Khaliullin2001}. The magnetic anisotropy that arises from this mechanism is an example of so-called order-from-disorder phenomena, where the fluctuation lifts the accidental degeneracy. This is a distinct phenomenon from the spins selecting a particular direction as a consequence of the exchange anisotropy.

In the high temperature paramagnetic phase, on the other hand, the directional-dependent interaction induce magnetic anisotrpy. In a recent paper, the author and his collaborator reported an unusual temperature dependence of nonlinear magnetic susceptibility in a frustrated fcc magnet with the directional-dependent interactions~\cite{Ishizuka2015_2}, which the direction of magnetic anisotropy changes between the ground state and the high-temperature paramagnetic phase. This is a separate phenomenon from the temperature-dependent magnetic anisotropy in lanthanids ions. For lanthanides, the temerature dependence of the anisotropy is attributed to the temperature dependence of the population of electrons in the excited orbital state of $f$ electrons~\cite{Bleaney1972,Mironov2002}. In contrast, in the mechanism proposed in Ref.~\onlinecite{Ishizuka2015_2}, the anisotropy arises from the anisotropic interactions, therefore, should be observed well below the temperature that corresponds to the orbital splitting. This situation is more likely for the heavy transition metal oxides, as the energy scale of spin-orbit interaction, Hubbard interaction between the electrons, and the crystal field splittings are roughly in the order of 0.1-1eV, well above the room temperature.

The result suggested that the interplay of bond-dependent interactions and fluctuations potentially give rise to rich physics, not only in the low-temperature ordered phase but also in the high-temperature paramagnetic phase. So far, however, the basic properties of such effect have not been studied. In this paper, we address this question by utilizing a high-temperature expansion. We first analyse different terms in the high-temperature expansion, and discuss how the abisotropy arise from the directional-dependent interaction. In the later half, we apply the result to archetypal models with the bond-dependent Ising type interactions: Heisenberg-Kitaev model on a honeycomb lattice and a cubic Heisenberg model with the directional-dependent interactions. By using the high temperature expansion up to second order in the inverse temperature, we study the external field dependence of the free energy. We show that the second order terms from the bond-dependent interactions give rise to nonlinear susceptibility, which shows maximum along $\langle100\rangle$ directions of the spin axis. We also discuss the possibiliy of observing the anisotropy by torque magnetometry. The result is potentially applicable to iridium oxides~\cite{Singh2010,Liu2011,Choi2012,Ye2012,Singh2012,Lovesey2012,Alpichshev2015} and RuCl$_3$~\cite{Plumb2014,Kim2015,Sears2015,Banerjee2016,Kubota2015,Johnson2015,Sandilands2015}.

The remainder of the paper is organized as follows. In Sec.~\ref{sec:HiT}, we briefly introduce the formalism we used for the high temperature expansion. Different terms that appear in the second order expansion is discussed in Sec.~\ref{sec:bond}. We discuss how and what kind of anisotropy arises from the different terms. In Secs.~\ref{sec:HK} and \ref{sec:cubic}, we apply the method to two different models, both being an effective model for different types of iridium oxides. Section~\ref{sec:HK} is devoted to the discussion on the Heisenberg-Kitaev model on the honeycomb lattice. In Sec.~\ref{sec:cubic} we apply the method to a cubic lattice model which is potentially an effective model for the perovskite iridates. Section~\ref{sec:summary} is devoted to discussion and summary.

\section{High Temperature Expansion} \label{sec:HiT}
     
In this paper, we study the magnetic anisotropy in the paramagnetic phase using high-temperature expansion. To be concrete, we consider a spin Hamiltonian: 
\begin{eqnarray}
  H&=&H_i+H_0\\
  H_i&=&-\frac12\sum_{i,j} J_{ij}^{\alpha\gamma}S_i^\alpha S_j^\gamma\\
  H_0&=&-h_\alpha\sum_i S_i^\alpha.
  \label{eq:hamil}
\end{eqnarray}
Here, $S_i^\alpha$ is the $\alpha=x,y,z$ component of the spin on $i$th site, and $J_{ij}^{\alpha\gamma}$ is the exchange coupling between spins on $i$th and $j$th site. $h_\alpha$ is the $\alpha$ component of external magnetic field. The free energy for this Hamiltonian is given by
\begin{eqnarray}
  \beta F&=&-\log Z,\\
  &=&-\sum_l \frac1{l!}\lambda_l-\log Z_0,
\end{eqnarray}
where $\beta=T^{-1}$ is the inverse temperature and $Z$ is the distribution function. In the second line, 
\begin{eqnarray}
  Z_0=\text{Tr} \exp\{-\beta H_0\},
  \label{eq:z0}
\end{eqnarray}
and $\lambda_i$ is the expansion of $i$th orderin $\beta$, which the first and second orders are given by
\begin{eqnarray}
  \lambda_1&=&\sum_{i\ge j}\left<S_i^\alpha \beta J_{ij}^{\alpha\gamma} S_j^\gamma\right>\\
  \lambda_2&=& \sum_{i\ge j}\left<\left(S_i^\alpha \beta J_{ij}^{\alpha\gamma} S_j^\gamma\right)^2\right> -\lambda_1^2.
  \label{eq:lambda}
\end{eqnarray}
Here, $\langle\cdots \rangle$ is the thermal average
\begin{eqnarray}
\langle\cdots\rangle = \frac1{Z_0}{\rm Tr}\left( \cdots e^{-\beta H_0}\right),
\end{eqnarray}
where $Z_0={\rm Tr}\left( e^{-\beta H_0}\right)$ is the distribution function for $H_0$. Unlike the high-temperature expansion for the Heisenberg model, where total $S_z$ is conserved in $H_i$, for arbitrary choice of the field direction, the general form of $\lambda_i$ become more complicated due to the non-comutativity of $H_0$ and $H_i$. However, up to the second order in $J/T$, the form of equation remains the same with that for the Heisenberg model, with $H_i$ replaced by the Hamiltonian of consideration~\cite{Ishizuka2015_2}. 

In the subsequent sections, we focus on a Hamiltonian with only the nearest-neighbor interactions of the form
\begin{eqnarray}
  J_{ij}^{\alpha\gamma} = J_H \delta_{\alpha\gamma}+J_K n_{ij}^\alpha n_{ij}^\gamma,
\end{eqnarray}
or with spin operators,
\begin{eqnarray}
  J_H \bm S_i \cdot \bm S_j+J_K (\bm S_i\cdot\bm n_{ij}) (\bm S_j\cdot\bm n_{ij}),
\end{eqnarray}
where $\bm S_i=(S_i^x,S_i^y,S_i^z)$. Here, the first term is the Heisenberg interaction, and the second term is the Ising-type directional-dependent interaction, which the Ising axis $\bm n_{ij}$ depends on bonds. In the presence of spin-orbit interaction, this interaction is generally allowed by the symmetry, and have also been derived microscopically~\cite{Jackeli2009}.

\section{Isolated Bond} \label{sec:bond}

As the main focus of this paper is on the effect of second order term in the high-temperature expansion, we here consider the contribution from each bond in the second order of $\beta$. The second order term, $\lambda_2$ consists of two different terms:
\begin{eqnarray}
  \lambda_2^{(1)}=\sum_{\alpha,\eta,\gamma,\delta} K_{ij}^{\alpha\eta}K_{ij}^{\gamma\delta}\left( \bar S_{\alpha\eta} \bar S_{\gamma\delta}-m_\alpha m_\eta m_\gamma m_\delta\right)
\end{eqnarray}
and
\begin{equation}
  \lambda_2^{(2)}=\sum_{\alpha,\eta,\gamma,\delta} (K_{ij}^{\alpha\eta}K_{jk}^{\gamma\delta}+K_{kj}^{\alpha\eta}K_{ji}^{\gamma\delta}) m_\alpha\left( \bar S_{\eta\gamma}-m_\eta m_\gamma\right) m_\delta.
\end{equation}
Here, $m_\alpha=\langle S^\alpha_i\rangle$ ($\alpha=x,y,z$) is the $\alpha$ component of the thermal average of the magnetization, $K_{ij}=\beta J_{ij}$, and
\begin{equation}
  \bar{S}_{\alpha\gamma}=\langle S_i^\alpha S_i^\gamma\rangle.
\end{equation}
We here ignored the site index as it does not depend on the site $i$. In the second order in the expansion, $\lambda_2^{(1)}$ involves only one bond, while $\lambda_2^{(2)}$ involves two bonds that shares a site ($j$ in the above case). In Sec.~\ref{sec:bond21}, we consider $\lambda_2^{(1)}$; $\lambda_2^{(2)}$ is discussed in Sec.~\ref{sec:bond22}.

\subsection{Second Order Expansion I: $\lambda_2^{(1)}$ Terms}\label{sec:bond21}

For a general $\bm n$, $\lambda_2^{(1)}$ reads
\begin{eqnarray}
  \lambda_2^{(1)}&=&-\left\{K_H m^2+K_K\left(\bm m\cdot \bm n\right)^2 \right\}^2 + K_H^2 \sum_{\alpha,\gamma} S_{\alpha\gamma}^2\nonumber\\
  && + K_HK_K \sum_{\alpha,\delta,\gamma} \left\{n_\alpha S_{\alpha\gamma} n_\delta S_{\delta\gamma}+n_\alpha S_{\gamma\alpha} n_\delta S_{\gamma\delta}\right\}\nonumber\\
  &&+ K_K^2 \left( \sum_{\alpha,\delta} n_\alpha S_{\alpha\delta} n_\delta\right)^2\label{eq:lambda21-S}
\end{eqnarray}
where $K_H=\beta J_H$, $K_K=\beta J_K$, and $m=|\bm m|$ is the average size of moment with
\begin{eqnarray}
  \bm m=\left(m_x,m_y,m_z\right).
\end{eqnarray}
Using quadrupolar parameters,
\begin{eqnarray}
Q_{xy}&=&\bar S_{xy}+\bar S_{yx},\\
Q_{yz}&=&\bar S_{yz}+\bar S_{zy},\\
Q_{zx}&=&\bar S_{zx}+\bar S_{xz},\\
Q_{x^2-y^2}&=&\bar S_{xx}-\bar S_{yy},\\
Q_{3z^2-r^2}&=&\frac1{\sqrt3}\left\{3\bar S_{zz}-S(S+1)\right\},
  \label{eq:Qparams}
\end{eqnarray}
$\lambda_2^{(1)}$ reads
\begin{widetext}
\begin{eqnarray}
  \lambda_2^{(1)}&=&-\left\{K_H m^2+K_K\left(\bm m\cdot \bm n\right)^2 \right\}^2 + K_H^2 \left\{\frac13S^2(S+1)^2+\frac12R-\frac12m^2\right\}+ K_K^2 \left\{\frac13S(S+1)+\frac12R_{nn}\right\}^2 \nonumber\\
  &&+2K_HK_K\Re\left[\left\{n_x\left(\frac23S(S+1)+Q_{x^2-y^2}-\frac1{\sqrt3}Q_{z^2-r^2}\right)+n_y \frac{Q_{xy}+i m_z}2+n_z\frac{Q_{zx}+i m_y}2\right\}^2\right]\nonumber\\
  &&+2K_HK_K\Re\left[\left\{n_x \frac{Q_{xy}-i m_z}2+n_y\left(\frac23S(S+1)-Q_{x^2-y^2}-\frac1{\sqrt3}Q_{z^2-r^2}\right)+n_z \frac{Q_{yz}+i m_x}2\right\}^2\right]\nonumber\\
  &&+2K_HK_K\Re\left[\left\{n_x\frac{Q_{zx}-i m_y}2+n_y\frac{Q_{yz}-i m_x}2+n_z\left(\frac23S(S+1)+\frac2{\sqrt3}Q_{z^2-r^2}\right)\right\}^2\right],\label{eq:lambda21-Q}
\end{eqnarray}
\end{widetext}
where
\begin{eqnarray}
  R= Q_{xy}^2 + Q_{yz}^2 + Q_{zx}^2 + Q_{x^2-y^2}^2 + Q_{3z^2-r^2}^2,
\end{eqnarray}
and
\begin{eqnarray}
  R_{ab}&=& Q_{xy}^{(ab)} Q_{xy} + Q_{yz}^{(ab)} Q_{yz} + Q_{zx}^{(ab)} Q_{zx}\nonumber\\
  &&\qquad+ Q_{x^2-y^2}^{(ab)} Q_{x^2-y^2} + Q_{3z^2-r^2}^{(ab)} Q_{3z^2-r^2}.
\end{eqnarray}
with 
\begin{eqnarray}
  Q^{(ab)}_{xy}&=& a_xb_y+a_yb_x\\
  Q^{(ab)}_{yz}&=& a_yb_z+a_zb_y\\
  Q^{(ab)}_{zx}&=& a_xb_z+a_zb_x\\
  Q^{(ab)}_{x^2-y^2}&=& a_xb_x-a_yb_y\\
  Q^{(ab)}_{3z^2-r^2}&=& \frac1{\sqrt3}\left(2a_zb_z-a_xb_x-a_yb_y\right).
\end{eqnarray}
Here, we used the commutation relation of spin operators,
\begin{equation}
  \left[S_i^\alpha,S_i^\beta\right]=i\sum_\gamma\epsilon_{\alpha\beta\gamma}S_i^\gamma,
\end{equation}
where $\epsilon_{\alpha\beta\gamma}$ is the antisymmetric tensor.

If we take $\bm n=(0,0,1)$, Eq.~\ref{eq:lambda21-Q} reads
\begin{eqnarray}
  \lambda_2^{(1)}&=& \frac{K_H^2}3S^2(S+1)^2-K_H^2m^4 - \frac{K_H(K_H+K_K)}2m^2\nonumber\\
  &&+ \frac{K_HK_K}2m_z^2 - 2K_HK_K m^2m_z^2 - K_K^2 m_z^4\nonumber\\
  &&+ \frac{K_H^2}2R + \frac{K_HK_K}2\left(Q_{zx}^2+Q_{yz}^2\right)\nonumber\\
  &&+ \frac{K_K(2K_H+K_K)}3 \left(Q_{3z^2-r^2}+\frac{S(S+1)}{\sqrt3}\right)^2,\label{eq:lambda21-nz}
\end{eqnarray}
In $\lambda_2^{(1)}$, the magnetic anisotropy in the order of $O(m^2)$ is given by $K_HK_Km_z^2$ and the term in the last line in Eq.~\ref{eq:lambda21-nz}, favoring spins to point along the bond [the free energy is proportional to $-k_B T \lambda_2^{(1)}$]. However, as we see in the subsequent sections, the anisotropy in the order of $O(m^2)$ often cancels out due to the symmetry of the lattice. In such cases, contribution from $O(m^4)$ become the leading order in anisotropy. In Eq.~\ref{eq:lambda21-nz}, the contribution comes from $-K_HK_Km_z^2 m^2 -2K_K^2m_z^4$ and the quadrupolar operators in the third and fourth lines in Eq.~\ref{eq:lambda21-nz}, if all quadrupolar parameters are zero at $\bm h=(h^x,h^y,h^z)=\bm 0$. On the other hand, if the quadrupolar parameter remains nonzero, e.g., by crystal field effect, the anisotropic interactions may contributes to the magnetic anisotropy at a lower order of $m$.

In the case of $S=1/2$, which we will mainly consider in the subsequent sections, the quadrupolar operators vanish. In this case, Eq.~\ref{eq:lambda21-nz} is simplified to
\begin{eqnarray}
  \lambda_2^{(1)} &=&\frac{2J_H^2+(J_H+J_K)^2}{16}-\frac{J_H(J_H+J_K)}2m^2\nonumber\\
  &&+\frac{J_HJ_K}2m_z^2-J_H^2 m^4-2J_HJ_K m^2m_z^2-J_K^2m_z^4.\nonumber\\
\end{eqnarray}
In this equation, the last two terms contribute to $O(m^4)$ magnetic anisotropy. Among these two terms, $-2J_HJ_K m^2m_z^2$ essentially renormalize the uniaxial anisotropy. On the other hand, $-J_K^2m_z^4$ term gives the cubic anisotropy; this term prefers to point the spin perpendicular to $\bm n$. Therefore, even for $S=1/2$ case, the directional-dependent exchange interactions gives rise to $m_z^4$ magnetic anisotropy. This is a distinct feature from the magnetic anisotropy due to the crystal field, which magnetic anisotropy does not appear for $S=1/2$ except for the anisotropy in $g$ factors.

\subsection{Second Order Expansion II: $\lambda_2^{(2)}$ Terms}\label{sec:bond22}

We next consider $\lambda_2^{(2)}$. We here focus on the two bonds $(i,j)$ and $(j,k)$ that shares site $j$; we take $\bm n_{ij}=\bm n$ and $\hat{n}_{jk}=\bm n'$. For general $S$, $\lambda_2^{(2)}$ reads
\begin{eqnarray}
  \lambda_2^{(2)}&=&-2\left\{K_H m^2+K_K\left(\bm m\cdot \bm n\right)^2 \right\}\nonumber\\
  &&\qquad\qquad\times\left\{K_H m^2+K_K\left(\bm m\cdot \bm n'\right)^2\right\}\nonumber\\
  &&+\frac23S(S+1)\left[K_H^2 m^2+K_K^2(\bm m\cdot \bm n)(\bm m\cdot \bm n')(\bm n\cdot \bm n')\right]\nonumber\\
  &&+\frac23S(S+1)\left[K_HK_K(\bm m\cdot \bm n')^2+K_HK_K(\bm m\cdot \bm n)^2\right]\nonumber\\
  &&+K_H^2 R^{(mm)}+K_K^2(\bm m\cdot \bm n)(\bm m\cdot \bm n') R^{(nn')}\nonumber\\
  &&+K_HK_K(\bm m\cdot \bm n') R^{(mn')}+K_HK_K(\bm m\cdot \bm n) R^{(mn)}.\label{eq:lambda22-Q}
\end{eqnarray}
Similar to $\lambda_2^{(2)}$, the quadrupolar moments appears in the second order in the expansion. Therefore, when a quadrupolar moment appears, e.g., due to crystal fields, the quadrupoles contribute to the magnetic anisotropy in the quadratic order in the free energy, in addition to the single ion term. Experimentally, this is observed as the anisotropy in the magnetic susceptibility.

For $S=1/2$, as $Q=0$ for $Q=Q_{xy}$, $Q_{yz}$, $Q_{zx}$, $Q_{x^2-y^2}$, and $Q_{3z^2-r^2}$, Eq.~\ref{eq:lambda22-Q} simplifies as 
\begin{eqnarray}
  \lambda_2^{(2)} &=& -2\left\{K_H m^2+K_K\left(\bm m\cdot \bm n\right)^2 \right\}\nonumber\\
  &&\qquad\qquad\times\left\{K_H m^2+K_K\left(\bm m\cdot \bm n'\right)^2\right\}\nonumber\\
  &&+\frac12\left[K_H^2 m^2+K_K^2(\bm m\cdot \bm n)(\bm m\cdot \bm n')(\bm n\cdot \bm n')\right]\nonumber\\
  &&+\frac12\left[K_HK_K(\bm m\cdot \bm n')^2+K_HK_K(\bm m\cdot \bm n)^2\right].\label{eq:lambda22-Q2}
\end{eqnarray}
In $\lambda_2^{(1)}$, the magnetic anisotropy of order $O(m^2)$ arises from the terms in the second and the third lines in Eq.~\ref{eq:lambda22-Q2}. In many cases, however, these $O(m^2)$ terms cancel out due to the lattice symmetry, similar to the $O(m^2)$ terms in $\lambda_2^{(1)}$. For such cases, the leading order in the magnetic anisotropy arises from the $O(m^4)$ terms. These terms consists of
\begin{eqnarray}
  -2K_HK_K m^2\left(\bm m\cdot \bm n\right)^2-2K_HK_K m^2\left(\bm m\cdot \bm n'\right)^2,
\label{eq:lambda22Q-m41}
\end{eqnarray}
and
\begin{eqnarray}
  -2K_K^2\left(\bm m\cdot \bm n\right)^2\left(\bm m\cdot \bm n'\right)^2.
\label{eq:lambda22Q-m42}
\end{eqnarray}
We note that, similar to the case of $\lambda_2^{(1)}$, $\lambda_2^{(2)}$ also contributes to the $O(m^4)$ magnetic anisotropy, even for $S=1/2$. The former term gives additional contribution to the uniaxial anisotropy. Therefore, these terms also vansh when the uniaxial anisotropy is prohibited by symmetry.

Hence, we focus on the term in Eq.~\ref{eq:lambda22Q-m42}. As this term is equal to or smaller than zero for arbitrary choice of $\bm m$, the maximum of $\lambda_2^{(2)}$, hence the minimum in the free energy, is given by the direction of $\bm m$ that gives zero. From Eq.~\ref{eq:lambda22Q-m42}, it is obvious that Eq.~\ref{eq:lambda22Q-m42} become zero if $\bm m$ is perpendicular to one of the two vectors, $\bm n$ or $\bm n'$. Therefore, the bond prefers to point the magnetic moment perpendicular to one of the bonds.

\section{Heisenberg-Kitaev Model}\label{sec:HK}

In this section we consider a honeycomb lattice model with all $\bm n_{ij}$ being a unit vector along $x$, $y$, or $z$ depending on the bond direction [Fig.~\ref{fig:model}(a)]. This model is a pedagogical model with all angles between $\bm n_{ij}$ and $\bm n_{jk}$ being $\pi/2$ for the nearest-neighbor bonds that share a site. Nevertheless, this model have been proposed to be an effective model for honeycomb $4d$, $5d$ oxides~\cite{Jackeli2009}. By the high-temperature expansion method used in previous sections, we show how the anisotropy arises from the directional-dependent interactions.

\subsection{Model}

\begin{figure}
  \includegraphics[width=\linewidth]{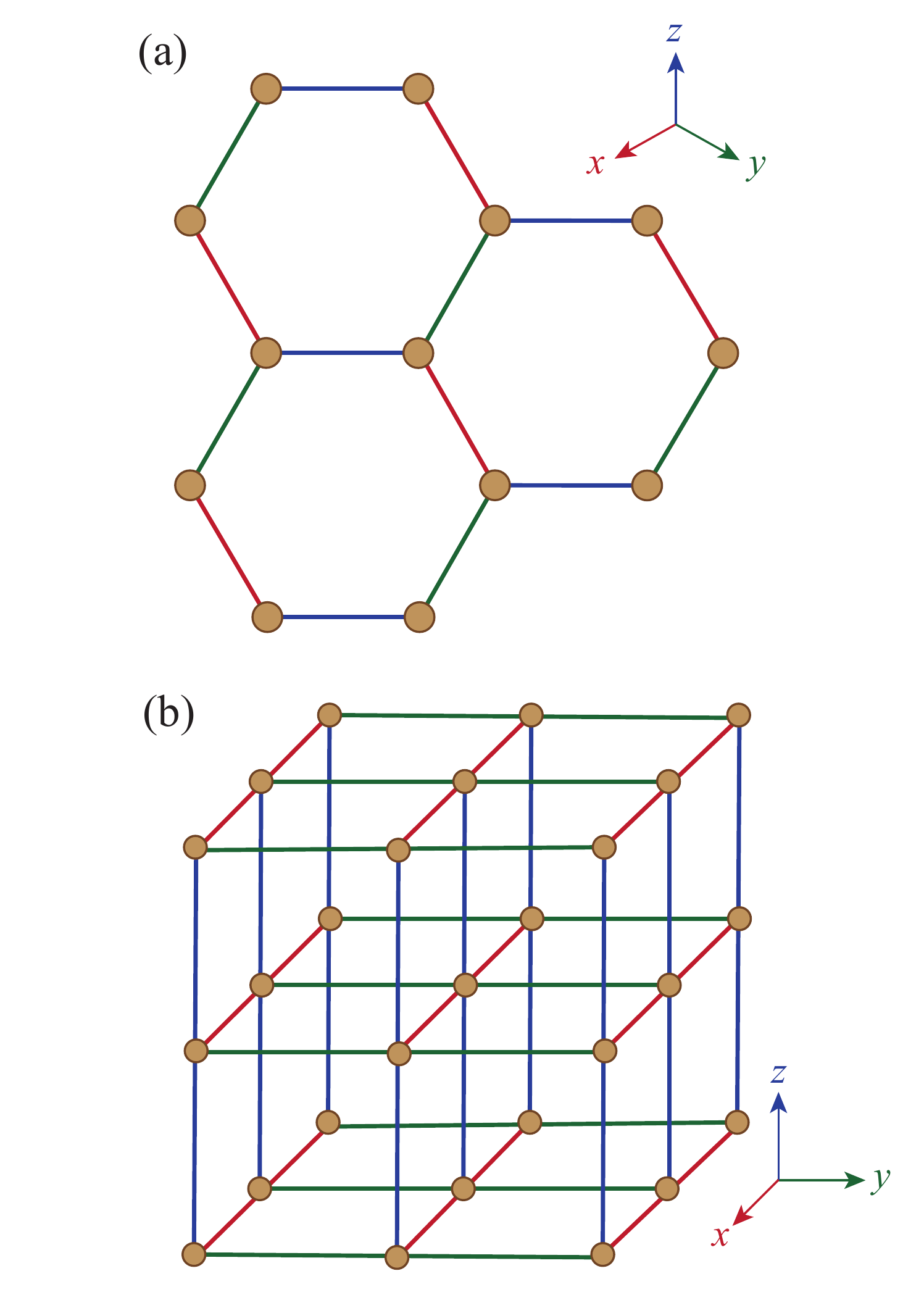}
  \caption{Schematic figure of the (a) Heisenberg-Kitaev and (b) cubic lattice model. The bonds with different colors corresponds to $J_K$ interactions for $\gamma=x$ (red), $y$ (green), and $z$ (blue). The arrows show the coordinate for the spins.}
  \label{fig:model}
\end{figure}

The Hamiltonian we consider in this section is given by
\begin{eqnarray}
H_{HK} = -J_H\sum_{\langle i,j\rangle} \bm S_i\cdot \bm S_j-J_K\sum_\gamma\sum_{\langle i,j\rangle_\gamma} S_i^\gamma S_j^\gamma.
\end{eqnarray}
Here, the first term is the nearest-neighbor Heisenberg interaction of spins $\bm{S}_i$ and $\bm{S}_j$; the sum is over all nearest-neighbor bonds on the honeycomb lattice. The second term is the directional-dependent interaction that depends on the direction of the bonds as shown in Fig.~\ref{fig:model}(a); the sum is over nearest-neighbor $\gamma$ bonds with $\gamma=x$, $y$, and $z$. This model has been proposed to be an effective model for the honeycomb Ir oxides with Ir$^{4+}$ ions~\cite{Jackeli2009,Chaloupka2010}, and its phase diagram was studied extensively~\cite{Chaloupka2010,Jiang2011,Reuther2011,Trousselet2011,Schaffer2012,Price2012,Price2013,Rau2014}.

\subsection{High-Temperature Expansion}

To study the magnetic anisotropy in the high-temperature paramagnetic phase, we here evaluate the free energy of the system using a high-temperature expansion. For the Heisenberg-Kitaev model, the first order in the expansion is given by
\begin{eqnarray}
  \lambda_1 &=&\left(3K_H+K_K\right)S^2 B_S^2(SL),
  \label{eq:lambdaHK1}
\end{eqnarray}
where $N$ is the number of unit cells, is the size of the spin, $B_S(x)$ is the Brillouin function, and $L=\beta g\mu h$ is the magnetic field renormalized by the inverse of the temperature $\beta$, where $g$ is the Lande's $g$ factor, and $\mu$ is the size of the magnetic moment. This term does not depend on the direction of the external field.

On the other hand, the second order in the expansion is given by
\begin{eqnarray}
  \lambda_2&=&\left(3K_H^2+2K_H K_K\right)\sum_{\alpha,\gamma} \left(\bar{S}_2^{\alpha\gamma}+4m^\alpha \bar{S}_1^{\alpha\gamma} m^\gamma\right)\nonumber\\
  &&+ K_K^2 \left(2\sum_{\alpha,\gamma}m^\alpha \bar{S}_1^{\alpha\gamma} m^\gamma + \sum_\alpha \bar{S}_2^{\alpha\alpha}-2m^\alpha \bar{S}_1^{\alpha\alpha} m^\alpha \right),\nonumber\\
  \label{eq:lambda2a}
\end{eqnarray}
where
\begin{eqnarray}
  \bar{S}_n^{\alpha\gamma} &=& \bar S_{\alpha\gamma}^n-m_\alpha^n m_\gamma^n.
  \label{eq:defSag}
\end{eqnarray}
In Eq.~\ref{eq:lambda2a}, the first term is rotationally symmetric, while the cubic anisotropy arises from the second term.

For $S=1/2$, the second term in Eq.~\ref{eq:lambda2a} reads
\begin{eqnarray}
  K_K^2\left( \frac3{16}-\frac18B_{1/2}^2(L/2) + \frac1{16} B_{1/2}^4(L/2)\sum_\alpha \hat{h}_\alpha^4 \right),
  \label{eq:lambda2d}
\end{eqnarray}
where $\hat{h}_\alpha$ is the $\alpha$th ($\alpha=x,y,z$) element of the unit vector along the external magnetic field $\bm h=(h_x,h_y,h_z)$. Therefore the leading order in $\beta$ of the anisotropy in the free energy is given by
\begin{eqnarray}
  f^{(ani)}_{2,0} =-\frac{J_K^2}{32T} B_{1/2}^4(L/2)\sum_\alpha \hat{h}_\alpha^4. \label{eq:faniHK}
\end{eqnarray}
Hence, the directional dependent interaction induces a temperature dependent magnetic anisotropy. For the Heisenberg-Kitaev model, the anisotropy favors $\langle100\rangle$ directions, i.e., along the direction of the $\bm n$. At a glance, the result in Eq.~\ref{eq:faniHK} appears to be contradicting with the argument in Sec.~\ref{sec:bond}, where the bond favors to point the magnetic moment perpendicular to $\bm n$. However, in the Heisenberg-Kitaev model, $\bm n$ for the three different bonds are perpendicular to each other. Therefore, the moment points along $\bm n$ for one of the bonds to make itself perpendicular to the $\bm n$ for other two bonds. We can directly see this by calculating the contribution from each bonds; the anisotropy term from $\lambda_2^{(2)}$ takes key role in selecting the direction of magnetic anisotropy. We note that, although the mechanism is different, this is the same direction that the spins prefers to point in the ground state~\cite{Chaloupka2010}. This behavior is different from what is reported for a fcc lattice model, which the direction of anisotropy changes for the paramagnetic and ordered phases~\cite{Ishizuka2015_2}.

By expanding the Brillouin function around $L=0$, and substitution $L=g\mu h/T$, to the leading order in $1/T$, $f_{HK}^{(ani)}$ reads
\begin{eqnarray}
  f^{(ani)}_{2,0} \sim -\frac{J_K^2}{32T} \sum_\alpha \left(\frac{g\mu h_\alpha}{2T}\right)^4.
\end{eqnarray}
Therefore, the leading order in the anisotropy appears $f_{HK}^{(ani)}\propto T^{-5}$. An interesting feature of the Eq.~\ref{eq:faniHK} is that the anisotropy term, in the leading order of $\beta$ only involves $J_K$, not $J_H$. This indicates that, for sufficiently high temperatures above the magnetic transition temperature, the anisotropy can simply be related to the anisotropic interaction. Therefore, observation of the magnetic anisotropy is potentially a method for evaluating the magnitude of the anisotropic interaction. 

\subsection{Lattice Distortion}

So far, we considered the symmetric Heisenberg-Kitaev model, which corresponds to the honeycomb network of ideal Ir$^{4+}$ octahedra~\cite{Jackeli2009}. For most of the candidate materials, however, the lattice is distorted\cite{Choi2012,Ye2012,Singh2012,OMalley2008,Johnson2015}, which modifies the interaction. This distortion is likely to induce an uniaxial anisotropy along one of the spin axis, as the magnitude of anisotropic interaction changes. To study how the lattice distortion affects the magnetic anisotropy, we consider the case in which the exchange interaction for $z$ bonds are enhanced/suppressed by $\Delta$,
\begin{eqnarray}
  H_{HK} &=& -J_H\sum_\gamma\sum_{\langle i,j\rangle_\gamma} (1+\Delta\delta_{\gamma z})\vec{S}_i\cdot \vec{S}_j\nonumber\\
  &&\qquad-J_K\sum_\gamma\sum_{\langle i,j\rangle_\gamma} (1+\Delta\delta_{\gamma z}) S_i^\gamma S_j^\gamma,\label{eq:Hhk_ani}
\end{eqnarray}
where $\delta_{\alpha\gamma}$ is the Kronecker's delta. Here, $\Delta=0$ corresponds to the isotropic case we considered in the previous section. With $\Delta\ne 0$, we here show that the distortion gives rise to additional terms in the anisotropic part of the free energy,
\begin{eqnarray}
  f_{ani} = f^{(ani)}_{2,0} + \left(f^{(ani)}_{1,1}+f^{(ani)}_{2,1}\right) \Delta + f^{(ani)}_{2,2} \Delta^2.
  \label{eq:faniHKgen}
\end{eqnarray}
Here, $f^{(ani)}_{2,0}$ is the anisotropy term given by Eq.~\ref{eq:faniHK}, which remains in the symmetric case ($\Delta=0$). To the linear order in $\Delta$, there are two terms, $f^{(ani)}_{1,1}$ and $f^{(ani)}_{2,1}$, that arises from the first order and second order in the expansion, respectively. On the other hand, $f^{(ani)}_{2,2}$ is the anisotropic term of order $O(\Delta^2)$; this term arise from the second order in expansion. We see that $f^{(ani)}_{1,1}$, $f^{(ani)}_{2,1}$, and $f^{(ani)}_{2,2}$ give rise to the uniaxial anisotropy along $z$ axis.

For the anisotropic model in Eq.~\ref{eq:Hhk_ani}, the first order in the high temperature expansion for $S=1/2$ reads
\begin{eqnarray}
  \lambda_1&=& N\left\{(3+\Delta)K_H+K_K\right\} m^2+ NK_K\Delta m_z^2,\\
  &=& \frac{N}4\left\{(3+\Delta)K_H+K_K\right\} B_S^2(L/2)\nonumber\\
  &&\quad \qquad \qquad \qquad + \frac{N}4K_K\Delta {\hat h}_z^2 B_S^2(L/2).\nonumber\\
\end{eqnarray}
Here, the first term gives the isotropic term that does not depend on the direction of the external field, while the second term gives uniaxial anisotropy about $z$ axis. Therefore, the free energy has an anisotropic term that comes from the second term,
\begin{eqnarray}
  f^{(ani)}_{1,1} &=& - J_K m_z^2.
\end{eqnarray}

With a similar analysis for the second order in the high-temperature expansion, in addition to the term given in Eq.~\ref{eq:faniHK}, we find anisotropic terms in the order of $\Delta$ and $\Delta^2$:
\begin{eqnarray}
  f^{(ani)}_{2,1} &=& \frac{2J_K(2J_H+J_K)}T m^2m_z^2 -\frac{J_K^2}T m_z^4,\\
  f^{(ani)}_{2,2} &=& \frac{J_HJ_K}{4T} \left(4m^2-1\right)m_z^2 + \frac{J_K^2}{2T}m_z^4.
\end{eqnarray}
These two terms also contribute to the uniaxial anisotropy for $z$ axis. Therefore, for $S=1/2$, the anisotropic part of the free energy reads
\begin{eqnarray}
  f_{ani} &=& c_1 m_z^2 + c_2 m_z^4 + c_3 \sum_\alpha m_\alpha^4
  \label{eq:faniHKfinal}
\end{eqnarray}
with
\begin{subequations}
  \begin{eqnarray}
  c_1 &=&-J_K\Delta\left[1+\Delta \frac{J_H}{4T}-m^2\left\{(4+\Delta)\frac{J_H}T+2\frac{J_K}T\right\}\right],\nonumber\\
   &\simeq&-\frac{J_K\Delta}{4}\left(1+\Delta \frac{J_H}T\right),\label{eq:faniHKc1} \\
  c_2 &=&-\frac{J_K^2}{2T}\Delta(2-\Delta),\\
  c_3 &=&-\frac{J_K^2}{4T}.
  \end{eqnarray}
  \label{eq:faniHKc}
\end{subequations}
Here, the third term in Eq.~\ref{eq:faniHKfinal} corresponds to Eq.~\ref{eq:faniHK} while the other two terms arise from the anisotropy as discussed in this section. In Eq.~\ref{eq:faniHKc1}, we ignored the second term in the brace as $m^2$ is expected to be small in the high-temperature paramagnetic phase.

\subsection{Torque Magnetometry}

\begin{figure}
  \includegraphics[width=\linewidth]{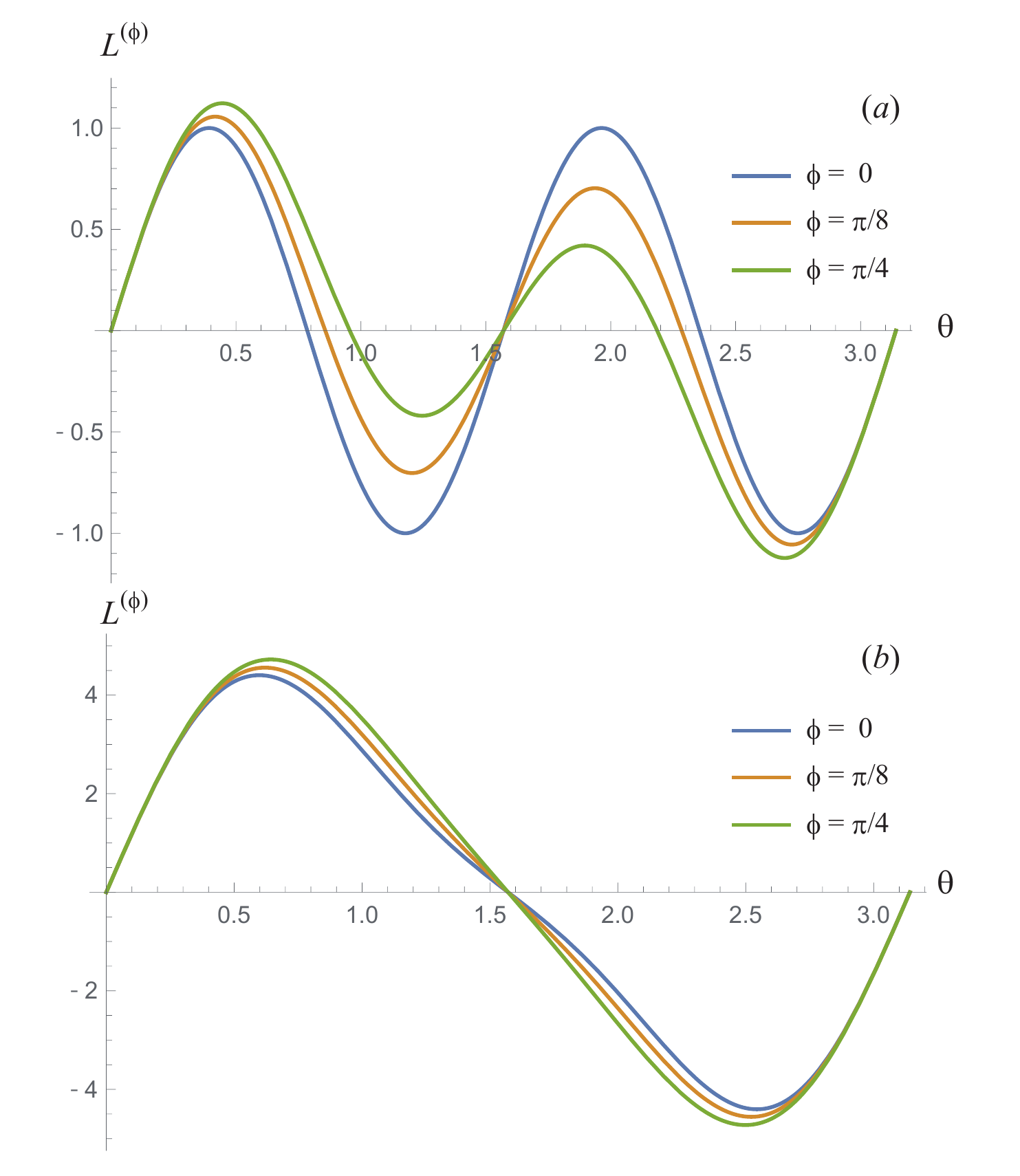}
  \caption{Magnetic torque curve $L\cdot \hat{\phi}$ calculated for angles $\phi=0,\pi/8,\pi/4$. The result is for (a) $c_1=c_2=0$ and $c_3=1$, and (b) $c_1=4$, $c_2=0$, and $c_3=1$.}
  \label{fig:torque}
\end{figure}

In the last, we discuss experimental measurement of the anisotropy by the magnetic torque induced by the anisotropic terms in Eq.~\ref{eq:faniHKfinal}. Using the free energy per unit cell, $f(\vec h)$, we calculate the magnetic torque by
\begin{eqnarray}
  \vec L(\vec h)&=& -\partial_\theta f(\vec h)\, \hat \phi + \frac1{\sin\theta}\partial_\phi f(\vec h)\,\hat \theta,\\
                     &=& -\partial_\theta f_{ani}(\vec h)\, \hat \phi + \frac1{\sin\theta}\partial_\phi f_{ani}(\vec h)\,\hat \theta,
\end{eqnarray}
where $(\theta,\phi)$ is the polar axis of the magnetic field, $\vec h=(\cos\phi\cos\theta,\cos\phi\sin\theta,\cos\theta)$, and $\hat \theta$ ($\hat \phi$) is the unit vector along $\theta$ ($\phi$) direction.

\begin{figure}
  \includegraphics[width=\linewidth]{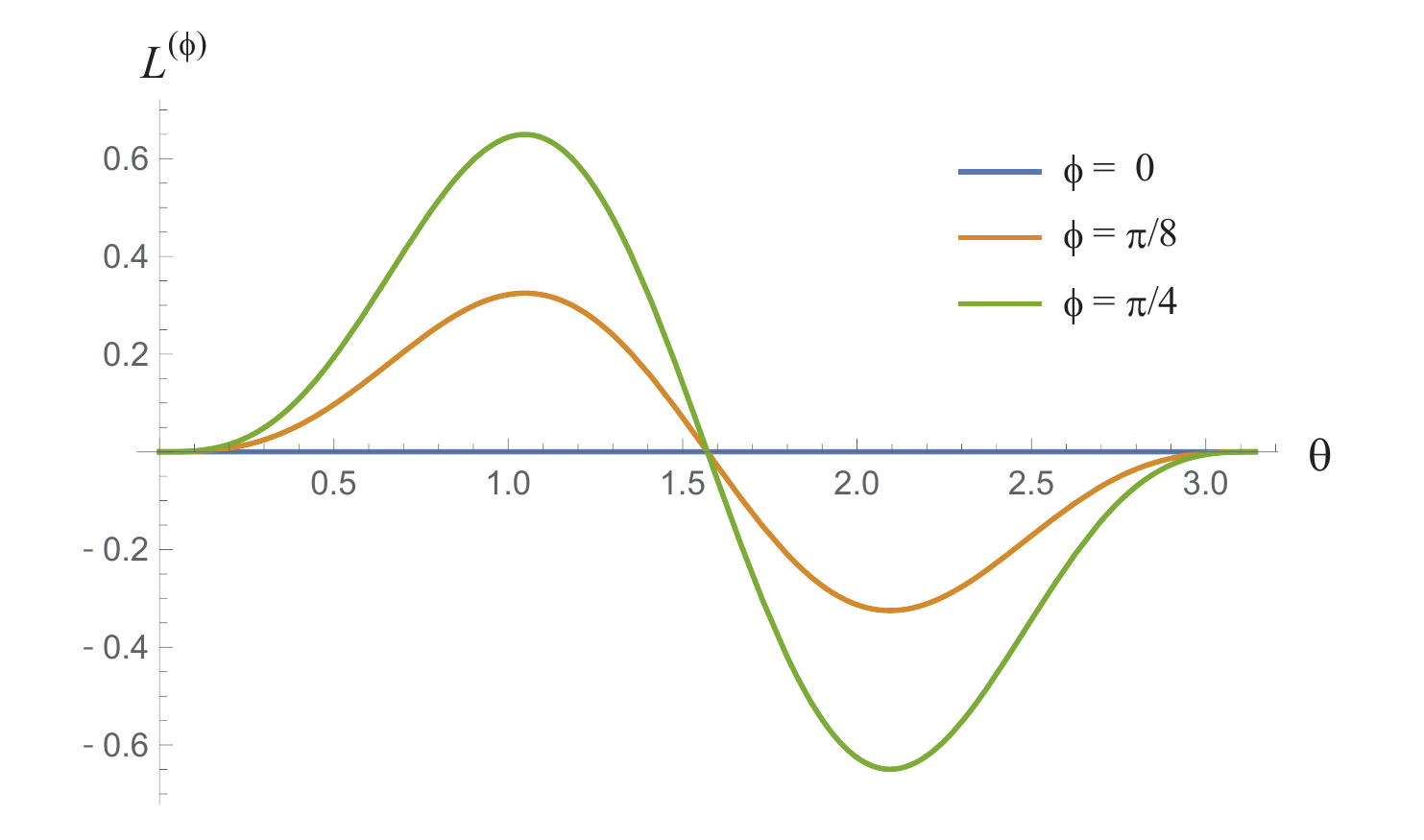}
  \caption{$\Delta L\cdot \hat{\phi}$ calculated for angles $\phi=0,\pi/8,\pi/4$. The result is for $c_3=1$.}
  \label{fig:dtorque}
\end{figure}

In Fig.~\ref{fig:torque}, we show the magnetic torque curve given by
\begin{equation}
  L^{(\phi)}(\theta,\phi)=\vec L(\vec h) \cdot \hat \phi=-\partial_\theta f_{ani}(\vec h).
\end{equation}
In Fig.~\ref{fig:torque}(a), we show the result for the symmetric Heisenberg-Kitaev model $\Delta=0$ (or $c_1=c_2=0$); the result shown in Fig.~\ref{fig:torque}(a) is for $c_3=1$. For this case, the magnetic torque arises only from $J_K$. Therefore, the measurement of magnetic anisotropy directly reflects the anisotropic interaction. In presence of the lattice distortion $\Delta$, however, the uniaxial anisotropy from the distortion also affects the magnetic torque. In this case, the contribution from $c_1$ expected to be much larger, as the $m_z^2$ term, $f_{1,1}^{(ani)}$, appear from the first order in the expansion. In Fig.~\ref{fig:torque}(b), we show results for $c_1=4$, $c_2=0$, and $c_3=1$. In this case, as given in Eq.~\ref{eq:faniHKc1}, $J_H$ also contributes to the magnetic anisotropy. Therefore, it reflects contribution from both $J_H$ and $J_K$. Similar change in the magnetic torque is also found when $c_2\ne 0$. Due to the difference in the symmetry, however, we find that the contribution from $J_K$ can be still isolated by taking the difference of magnetic torque for two different directions of $\phi$. For example, 
\begin{eqnarray}
  \Delta L^{(\phi)}(\theta,\phi)&=&L^{(\phi)}(\theta,\phi)-L^{(\phi)}(\theta,0),\\
                                     &=&2c_3\cos\theta\sin^3\theta\sin^2(2\phi),
\end{eqnarray}
only depends on $c_3$. The plot of $\Delta L^{(\phi)}_{HK}(\theta,\phi)$ for different $\phi$ is shown in Fig.~\ref{fig:dtorque} ($c_3=1$). As $c_3$ is the function of only $J_K$, not $J_H$, magnetic torque measurement potentially provides an experimental method to directly observe $J_K$.

Another important feature of the mangetic anisotropy that arise from $J_K$ is the temperature dependence. As $m=SB_S(SL)\propto h/T$ ($h\ll T$), the temperature dependence of the anisotropic term of free energy is $f_{2,0}^{(ani)} \propto T^{-5}$. Therefore, the magnetic torque from $c_3$ and $\Delta L^{(\phi)}(\theta,\phi)$ also increase $\propto T^{-5}$ with decreasing temperature, in the high temperature limit. The observation of the temperature dependence of the anisotropy potentially provides an method to isolate the anisotropy from $J_K$ from that of van Vleck susceptibility, which is temperature independent.

In the last, we provide an order estimate for $\Delta L^{(\phi)}(\theta,\phi)$. Assuming $T\sim 10$K, $H\sim 10$T, $J_K/k_B\sim 10$K, $\mu_\text{eff}=1-2$, and number of ions $N\sim10^{23}$, the estimated magnitude of $c_3$ is about $10^{-11}-10^{-10}$ N$\cdot$m.

\section{Cubic Lattice Model} \label{sec:cubic}

\subsection{Model}

In this section, we consider a spin model on a cubic lattice, which is similar to the Heisenberg-Kitaev model studied in Sec.~\ref{sec:HK} [see Fig.~\ref{fig:model}(b)]; the Hamiltonian is given by
\begin{eqnarray}
H_{c} = -J_H\sum_{\langle i,j\rangle} \vec{S}_i\cdot \vec{S}_j-J_K\sum_\gamma\sum_{\langle i,j\rangle_\gamma} S_i^\gamma S_j^\gamma.
\end{eqnarray}
The first sum in the Hamiltonian is over all nearest-neighbor bonds, and the second sum is over the nearest neighbor bonds along $\gamma=x,y,z$ direction. This model is potentially an effective model for the transition metal perovskites with strong spin-orbit interactions~\cite{Jackeli2009}. A difference from Heisenberg-Kitaev model is that the $\theta=\pi$ bonds appears in the $\lambda_2^{(1)}$ term.

\subsection{High-temperature expansion}

Similar to the case of Heisenberg-Kitaev model in Sec.~\ref{sec:HK}, the first order expansion, $\lambda_1$, is given by 
\begin{eqnarray}
  \lambda_1&=&NS^2 B_S^2(SL)\left(3J_H+J_K\right),
  \label{eq:cubic1}
\end{eqnarray}
and the second order in the expansion by
\begin{eqnarray}
	\lambda_2 &=& \left(3K_H^2+2K_HK_K\right)\sum_{\alpha,\gamma}\left(10m_\alpha \bar S^{\alpha\gamma}_1 m_\gamma + \bar S^{\alpha\gamma}_2\right)\nonumber\\
            &&+4K_K^2\sum_{\alpha,\gamma}m_\alpha \bar S^{\alpha\gamma}_1 m_\gamma + K_K^2\sum_{\alpha}\left(\bar S^{\alpha\alpha}_2-m_\alpha^2 \bar S^{\alpha\alpha}_1\right)\nonumber\\
\end{eqnarray}
From the direct calculation, we find that the anisotropy term for the cubic model also arises from the $K_H^2$ order, which is given in the same form as in Eq.~\ref{eq:lambda2d}. Hence, it favors the field direction parallel to the bonds. This is also the same direction with the ground state anisotropy, which the anisotropy comes from order from disorder mechanism~\cite{Khaliullin2001}. As the result for the cubic lattice is the same with that of the Heisenberg-Kitaev model in Sec.~\ref{sec:HK}, the magnetic torque also behaves in the same manner. 

\section{Discussions and Summary} \label{sec:summary}

In this paper, we studied the magnetic anisotropy in the high-temperature magnetic phase. We presented that, in presence of directional-dependent interactions, the interaction gives rise to  $O(m^4)$ order anisotropy. This term contributes to the anisotropy in a distinct way from the $O(m^2)$ term that can be intuitively understood. Indeed, the anisotropy survives in the case of the Heisenberg-Kitaev and the cubic models, which the $O(m^2)$ order anisotropy cancels due to the symmetry. Moreover, the anisotropy arise even for the $S=1/2$ magnets, as we have shown for the Heisenberg-Kitaev and the cubic models. In these two models, the result appears to resemble that of the order-from-disorder phenomena in the ordered phase, of which the fluctuation selects spins to point along one of the bond~\cite{Chaloupka2010,Khaliullin2001}. 

However, the origin of this anisotropy is more complicated. According to the analysis on the isolated bond, we find that the $O(m^4)$ term prefers the spins to point perpendicular to the bond; the preferred direction in the two models (parallel to one of the bond) is a consequence of pointing the spin perpendicular to the other two (four) bonds in the Heisenberg-Kitaev (cubic) model. This mechanism for the anisotropy may give rise to a non-trivial consequence when the angle between the bonds are not $\pi/2$ or $\pi$. Indeed, in a recent paper, the author and his collaborator proposed that a sign flip of the anisotropic term in the Landau free energy occurs with changing temperature in the case of a fcc antiferromagnet\cite{Ishizuka2015_2}. As the temperature dependent magnetic anisotropy reflects the microscopic details of the interaction, observation of this phenomena may be useful to experimentally detecting the existance of the anisotropic interactions.

In honeycomb iridates, other interactions such as symmetric off-diagonal exchange interactions and further-neighbor interactions may also affect the magnetism of the material~\cite{Bhattacharjee2012,Yamaji2014,Katukuri2014,Winter2016,Chaloupka2016,Ohkubo2016}. Considerations on these interactions are left for future studies.

\acknowledgements

The author is grateful to G. Jackeli, Y. Motome, and H. Takagi for fruitful discussions. This work was supported by JSPS Grant-in-Aid for Scientific Research (No. 16H06717) from MEXT, Japan.

\end{document}